\documentclass[pdftex,a4paper,12pt]{article}
\usepackage{amssymb,amsmath,amsfonts,ntheorem,nomencl,mathrsfs}

\usepackage[arrow, matrix, curve]{xy}
\usepackage[latin1]{inputenc}

\newcommand{\IC}{\mathbb{C}}
\newcommand{\IR}{\mathbb{R}}

\newcommand{\IFF}{\mathscr{F}}
\newcommand{\IAA}{\mathscr{A}}

\newcommand{\IN}{\mathbb{N}}

\newcommand{\Id}{{\rm d}}

\newcommand{\f}{\frac}
\newcommand{\nn}{\nonumber}

\newtheorem{theorem}{THEOREM}[section]

\newtheorem{Lemma}[theorem]{Lemma}

\newtheorem{Remark}[theorem]{Remark}

\newtheorem{Theorem}[theorem]{Theorem}

\newtheorem{Proposition}[theorem]{Proposition}

\newtheorem{Definition}[theorem]{Definition}

\begin{document}
\begin{titlepage}

\title{Multiplicative matrix-valued functionals and the continuity properties of semigroups correspondings to partial differential operators with matrix-valued coefficients}
  \author{Batu Güneysu\footnote{E-Mail: gueneysu@math.uni-bonn.de}\\
   Mathematisches Institut\\
   Universität Bonn\\
   }

\end{titlepage}

\maketitle

 \begin{abstract} We define and examine certain matrix-valued multiplicative functionals with local Kato potential terms and use probabilistic techniques to prove that the semigroups of the corresponding partial differential operators with matrix-valued coefficients are spatially continuous and have a jointly continuous integral kernel. These partial differential operators include Yang-Mills type Hamiltonians and Pauli type Hamiltonians, with ``electrical'' potentials that are elements of the matrix-valued local Kato class.
\end{abstract}

\section{Main results}\label{zhn}

Let $\IR^n$ and $\IC^d$ both be equipped with the corresponding Euclidean metric $\left\|\bullet\right\|$. The associated operator norm on $\mathrm{Mat}(\IC^d):=\mathrm{Mat}_{d\times d}(\IC)$ will be denoted with the same symbol. We will use the following notation for any 
\[
\alpha\in\Omega^1\left( \IR^n,\mathrm{Mat}(\IC^d)\right), 
\]
the smooth $1$-forms on $\IR^n$ with values in $\mathrm{Mat}(\IC^d)$: Any such $\alpha$ can uniquely be written as $\alpha=\sum^{n}_{j=1} \alpha_j \Id x^j$ with
\begin{align}
\alpha_j = (\alpha^{k}_{j,l})^{1\leq k\leq d}_{1\leq l\leq d} \in \mathrm{C}^{\infty}(\IR^n,\mathrm{Mat}(\IC^d)),\>\>j=1,\dots,n.\label{z66}
\end{align}
Let $\mathscr{U}(d)$ denote the skew-Hermitian elements of $\mathrm{Mat}(\IC^d)$, that is, $\mathscr{U}(d)$ is the Lie algebra corresponding to the unitary group $\mathrm{U}(d)$. In this paper, we shall be concerned with probabilistic methods for self-adjoint operators in $\mathrm{L}^2(\IR^n,\IC^d)$ that are formally given by the differential expression 
\begin{align}
 \tau(\alpha,V)= &-\f{1}{2}\Delta-\f{1}{2}  \sum^n_{j=1}\alpha_j^2-\f{1}{2}\sum^n_{j=1} (\partial_j\alpha_j)  - \sum^n_{j=1}\alpha_j \partial_j+V,\label{cax}
\end{align}
where $\alpha\in\Omega^1\left( \IR^n,\mathscr{U}(d)\right)$ and where $V:\IR^n\to\mathrm{Mat}(\IC^d)$ is a {\it potential}, that is, a measurable function with $V(x)=V^*(x)$ for almost every (a.e.) $x\in\IR^n$. If $d=1$, then one has $\alpha=\mathrm{i}\tilde{\alpha}$ for some real-valued $\tilde{\alpha}=\sum^n_{j=1}\tilde{\alpha}_j\Id x^j$, so that $\tau(\alpha,V)$ is nothing but the nonrelativistic Hamiltonian corresponding to a charged particle in the magnetic field $\tilde{\alpha}\in\Omega^1_{\IR}(\IR^n)$ and the electrical potential $V:\IR^n\to\IR$,
\[
\tau(\alpha,V)= -\f{1}{2}\Delta+\f{1}{2}\sum^n_{j=1}\tilde{\alpha}^2_j   -\f{\mathrm{i}}{2} \mathrm{div}(\tilde{\alpha})  - \mathrm{i}\sum^n_{j=1}\tilde{\alpha}_j \partial_j+V.
\]

The following conventions will be used for our probabilistic considerations: For any $x\in\IR^n$ we will denote the usual Wiener probability space with
\[
\mathscr{P}^x:=(\Omega,\IFF,(\IFF_t)_{t\geq 0},\mathbb{P}^x), 
\]
where $\Omega=\mathrm{C}([0,\infty),\IR^n)$ and where $\mathbb{P}^x$ stands for the Wiener measure on $(\Omega,\IFF)$ which is concentrated on the paths $\omega:[0,\infty)\to\IR^n$ with $\omega(0)=x$. The underlying $\sigma$-algebra $\IFF$ and the filtration $\IFF_*$ will be the ones corresponding to the canonical process 
\begin{align}
X:[0,\infty)\times \Omega\to\IR^n,\label{ggb}
\end{align}
where $\IFF_*$ will be made right-continuous and complete (locally complete, if Girsanov techniques are used; here we implicitely use the results of section 5.6 in \cite{Ha}), whenever necessary. We consider the process $X$ given by (\ref{ggb}) as a Brownian motion starting in $x$ under $\mathbb{P}^x$ and we will write ``$\underline{\Id}$'' for Stratonovic differentials, whereas It\^{o} differentials will be written as ``$\Id$''. \\
Fix $x\in\IR^n$ now. If $\alpha\in\Omega^1(\IR^n,\mathrm{Mat}(\IC^d))$ and $V:\IR^n\to\mathrm{Mat}(\IC^d)$ is such that
\begin{align}
\mathbb{P}^x\left\lbrace \int^t_0\left\|V\left(X_s\right)\right\| \Id s <\infty \right\rbrace=1\>\>\text{ for all $t>0$,}\label{bb11}
\end{align}
then the processes
\begin{align}
&A^{\alpha,V}:=\sum^n_{j=1}\int^{\bullet}_0 \alpha_j(X_s) \underline{\Id} X^{j}_s-\int^{\bullet}_0 V\left(X_s\right) \Id s:[0,\infty)\times\Omega\longrightarrow\mathrm{Mat}(\IC^d),\nn\\
&B^{\alpha,V}:=A^{\alpha,V}+\f{1}{2}[A^{\alpha,V},A^{\alpha,V}]:[0,\infty)\times\Omega\longrightarrow\mathrm{Mat}(\IC^d),\label{z55}
\end{align}
where
\[
[A^{\alpha,V},A^{\alpha,V}]^{j}_{k}:= \sum^d_{l=1} [(A^{\alpha,V})^j_l,(A^{\alpha,V})^l_k]\>\>\text{ for $j,k=1,\dots,d$}
\]
is the quadratic covariation, are continuous semi-martingales. For any $l\in\IN$ and $t\geq 0$ let the simplex $t\Delta_l$ be given by
\[ 
t\Delta_l:=\left.\Big\{ (t_1,\dots,t_{l})\right|0\leq t_1\leq\cdots\leq t_{l}\leq t\Big\}.
\]

Defining a stochastic path ordered exponential \footnote{This notation has to be understood as \[\mathscr{A}^{\alpha,V}_t =\mathbf{1}+B^{\alpha,V}_t+\int^t_0 B^{\alpha,V}_{s}\Id B^{\alpha,V}_{s}+\int^t_0 \left(  \int^{s}_0  B^{\alpha,V}_{r} \Id B^{\alpha,V}_{r} \right)      \Id B^{\alpha,V}_{s}+\cdots.\]} by
\begin{align}
\mathscr{A}^{\alpha,V}_t:= \mathbf{1}+\sum^{\infty}_{l=1} \int_{t\Delta_{l} } \Id
B^{\alpha,V}_{t_1}\cdots \Id B^{\alpha,V}_{t_{l}}, \label{for2}
\end{align}
where the convergence is $\mathbb{P}^x$-a.s. uniformly in compact subsets of $[0,\infty)$ \cite{pos}, one finds that 
\[
\mathscr{A}^{\alpha,V}:[0,\infty)\times\Omega\longrightarrow\mathrm{Mat}(\IC^d)
\]
is uniquely determined as the solution of
\begin{align}
 \mathscr{A}^{\alpha,V}_t = \mathbf{1}+\int^t_0 \mathscr{A}^{\alpha,V}_s \Id B^{\alpha,V}_s  \label{pop0}
\end{align}
under $\mathbb{P}^x$. It is easily seen that 
\begin{align}
 \mathscr{A}^{\alpha,V}_t &= \mathbf{1}+\int^t_0 \mathscr{A}^{\alpha,V}_s \underline{\Id} A^{\alpha,V}_s,\label{pop1}\\
  \mathscr{A}^{\alpha,V,*}_t &= \mathbf{1}+ \int^t_0  \left( \underline{\Id} A^{\alpha,V,*}_s\right)   \mathscr{A}^{\alpha,V,*}_s ,\label{pop2}\\
  \mathscr{A}^{\alpha,V,-1}_t &=\mathbf{1} -\int^t_0  (\underline{\Id} A^{\alpha,V}_s)  \mathscr{A}^{\alpha,V,-1}_s.\label{pop3}
\end{align}

\begin{Remark} If $d=1$ and $\alpha=\mathrm{i}\tilde{\alpha}$ for some $\tilde{\alpha}\in\Omega^1_{\IR}(\IR^n)$, then one easily finds 
\begin{align}
 \mathscr{A}^{\alpha,V}&=\exp\left( \mathrm{i}\sum^n_{j=1}\int^{\bullet}_0 \tilde{\alpha}_j(X_s) \underline{\Id} X^{j}_s-\int^{\bullet}_0 V\left(X_s\right) \Id s\right)\nn\\
&=\exp\left( \mathrm{i}\sum^n_{j=1}\int^{\bullet}_0 \tilde{\alpha}_j(X_s) \Id X^{j}_s+\f{\mathrm{i}}{2}\int^{\bullet}_0\mathrm{div}(\tilde{\alpha})(X_s)\Id s-\int^{\bullet}_0 V\left(X_s\right) \Id s\right),\label{zzay}
\end{align}
an expression which is well-known from the classical Feynman-Kac-It\^{o} formula. In particular, the identity
\[
A^{\alpha,V}_t(\omega(s+\bullet))=A^{\alpha,V}_{s+t}(\omega)-A^{\alpha,V}_s(\omega) \>\>\text{ for all $s,t\geq 0$ }
\]
(which follows approximating the integrals in the definition of $A^{\alpha,V}$ with Riemann type sums and $\mathrm{e}^{z_1+z_2}=\mathrm{e}^{z_1}\mathrm{e}^{z_2}$ directly imply the following relation for all $s,t\geq 0$,
\begin{align}
\mathscr{A}^{\alpha,V}_{s+t}(\omega)=\mathscr{A}^{\alpha,V}_{s}(\omega)\mathscr{A}^{\alpha,V}_{t}(\omega(s+\bullet))\>\>\text{ for $\mathbb{P}^x$-a.e. $\omega\in\Omega$}.\label{mmb}
\end{align}

\end{Remark}

Although one does not have such an explicit expression for $\mathscr{A}^{\alpha,V}$ for $d>1$, one can still prove the multiplicative property (\ref{mmb}) in the general case:

\begin{Lemma}\label{tcy} The process $\mathscr{A}^{\alpha,V}$ is a multiplicative matrix-valued functional, that is, for any $s,t \geq 0$ one has
\begin{align}
\mathscr{A}^{\alpha,V}_{s+t}=\mathscr{A}^{\alpha,V}_{s}\left(\mathscr{A}^{\alpha,V}_{t}\circ \vartheta_s\right)\>\>\text{ $\mathbb{P}^x$-a.s.},\label{aaaq}
\end{align}
where $\vartheta_s(\omega)=\omega(s+\bullet)$ stands for the shift operator on $\Omega$.  
\end{Lemma}

{\it Proof.} We fix $s$ and define $\IAA:=\IAA^{\alpha,V}$ and $A:=A^{\alpha,V}$. The following stochastic integrals are all understood with respect to $\mathbb{P}^x$. We will prove that the processes $ \mathscr{A}_{s+\bullet}$ and $ \mathscr{A}_{s}(\mathscr{A}_{\bullet}\circ \vartheta_s)$ both solve the following Stratonovic initial value problem (with respect to the filtration $(\IFF_{s+t})_{t\geq 0}$): 
\begin{align}
U_t= \mathscr{A}_{s} + \int^t_0 U_r \underline{\Id}_r A_{s+r}.\label{vv11}
\end{align}
To this end, note that (\ref{pop1}) directly implies
\begin{align}
\IAA_{s+t}&=\mathbf{1}+\int^{s+t}_0 \IAA_r  \underline{\Id}_r A_{r}=\mathbf{1}+\int^{s}_0 \IAA_r  \underline{\Id}_r A_{r}+\int^{t}_0 \IAA_{r+s}  \underline{\Id}_r A_{r+s}\nn\\
&=\mathscr{A}_{s} + \int^t_0 \IAA_{s+r} \underline{\Id}_r A_{s+r}.
\end{align}
On the other hand, the identity
\[
A_r\circ\vartheta_s=A_{s+r}-A_s\>\>\text{ $\mathbb{P}^x$-a.s. for all $r\geq 0$} 
\]
implies the second identity in
\begin{align}
\mathscr{A}_{t}\circ \vartheta_s &= \mathbf{1}+\left(\int^{t}_0 \IAA_r  \underline{\Id}_r A_{r}\right)\circ \vartheta_s\nn\\
&= \mathbf{1}+\int^{t}_0 \IAA_r\circ \vartheta_s \  \underline{\Id}_r A_{r+s},
\end{align}
so that the desired equality follows from multiplying the latter equation with $\IAA_s$ from the left. \vspace{0.5mm}

\hfill$\blacksquare$\vspace{2mm}

We refer the reader to \cite{pins} and the references therein for a detailed study of multiplicative matrix-valued functionals. \\
Matrix-valued Kato functions can be defined as follows:

\begin{Definition} A measurable function $V:\IR^n\to\mathrm{Mat}(\IC^d)$ is said to belong to the {\it $\mathrm{Mat}(\IC^d)$-valued Kato class of $\IR^n$}, if one has
\[
\lim_{t\searrow 0}\sup_{x\in \IR^n} \mathbb{E}^x\left[\int^t_0
\left\|V(X_s)\right\|\Id s\right]=0,
\]
and $V$ is said to be in the {\it $\mathrm{Mat}(\IC^d)$-valued local Kato class of $\IR^n$}, if $1_K V$ is in the corresponding Kato class for any compact subset $K\subset M$. 
\end{Definition}

We write $\mathcal{K}(\IR^n,\mathrm{Mat}(\IC^d))$ and $\mathcal{K}_{\mathrm{loc}}(\IR^n,\mathrm{Mat}(\IC^d))$ for the Kato and the local Kato class, respectively. Note also that for a measurable function $V:\IR^n\to\mathrm{Mat}(\IC^d)$ the condition $V\in\mathcal{K}_{\mathrm{loc}}(\IR^n,\mathrm{Mat}(\IC^d))$ is equivalent to 
\[
\varphi V \in \mathcal{K}(\IR^n,\mathrm{Mat}(\IC^d))\>\>\text{ for any $\varphi\in\mathrm{C}^{\infty}_0(\IR^n)$}.
\]

For any $p$ such that $p\geq 1$ if $m=1$, and $p > m/2$ if $m\geq 2$, one has
\begin{align}
\mathrm{L}^p_{\mathrm{loc}}(\IR^n,\mathrm{Mat}(\IC^d))\subset \mathcal{K}_{\mathrm{loc}}(\IR^n,\mathrm{Mat}(\IC^d))\subset\mathrm{L}^1_{\mathrm{loc}}(\IR^n,\mathrm{Mat}(\IC^d)).\label{tgb}
\end{align}

These inclusions may be found in \cite{Aiz}.

\begin{Remark}\label{bem} We will frequently use a simple consequence of the definition of the Kato class: If $V\in\mathcal{K}(\IR^n,\mathrm{Mat}(\IC^d))$, then the Chapman-Kolmogorov equation for the heat kernel of $\IR^n$ shows that for all $t\geq 0$, 
\begin{align}
\sup_{x\in \IR^n} \mathbb{E}^x\left[\int^t_0
\left\|V\left(X_s\right)\right\|\Id s\right]<\infty.\label{cvb}
\end{align}

Using this and the continuity of Brownian motion easily implies the following: If $V\in\mathcal{K}_{\mathrm{loc}}(\IR^n,\mathrm{Mat}(\IC^d))$, then
\begin{align}
\mathbb{P}^x\left\lbrace \int^t_0\left\|V\left(X_s\right)\right\| \Id s <\infty \right\rbrace=1. \label{ccx}
\end{align}
\end{Remark}

We can now prove two convergence results for $\IAA^{\alpha,V}$ that will turn out to be closely related to continuity properties of the semigroup that corresponds to an operator of the form $\tau(\alpha,V)$ as in (\ref{cax}). To this end, a potential $V$ will be called {\it nonnegative}, if all eigenvalues of the matrix $V(x):\IC^d\to\IC^d$ are nonnegative for a.e. $x\in \IR^n$. The following two lemmas extend lemma C.3 and lemma C.5 in \cite{Bro1} to the matrix-valued setting:

\begin{Proposition}\label{hiu} Let $V$ be a nonnegative potential with
\[
V\in\mathcal{K}(\IR^n,\mathrm{Mat}(\IC^d))
\]
and let $\alpha\in\Omega^1(\IR^n, \mathscr{U}(d))$ be such that 
\begin{align}
\max_{\overset{i=1,\dots,n}{j,k=1,\dots,d}}\left|\partial_i\alpha^j_{i,k}\right|\in\mathcal{K}(\IR^n),\>\max_{\overset{i=1,\dots,n}{j,k,l,m=1,\dots,d}}\left|\alpha^{j}_{i,l}\alpha^{k}_{i,m}\right|\in  \mathcal{K}(\IR^n), \label{ccr}
\end{align}
where the meaning of the indices in (\ref{ccr}) is as in (\ref{z66}). Then one has 
\begin{align}
 \lim_{t\searrow 0}\sup_{x\in\IR^n}
\mathbb{E}^x\left[\left\|\mathscr{A}^{\alpha,V}_{t}-\mathbf{1}\right\|\right]=0.\label{eop}
\end{align}
\end{Proposition}

\begin{Remark} If $d=1$, then the estimate $|\mathrm{e}^z-1|\leq C|z|\mathrm{e}^{\max\{\mathrm{Re}(z),0\}}$ for all $z\in\IC$ combined with (\ref{zzay}) and $V\geq 0$ directly imply
\begin{align}
&\mathbb{E}^x\left[\left|\mathscr{A}^{\alpha,V}_{t}-\mathbf{1}\right|\right]\nn\\
&\leq C \  \mathbb{E}^x\left[\left| \mathrm{i}\sum^n_{j=1}\int^{t}_0 \tilde{\alpha}_j(X_s) \Id X^{j}_s+\f{\mathrm{i}}{2}\int^t_0\mathrm{div}(\tilde{\alpha})(X_s)\Id s-\int^{t}_0 V\left(X_s\right) \Id s\right|\right],\label{dd2}
\end{align}
so that in this case (\ref{eop}) follows immediately from the It\^{o} isometry and the assumptions on $(\alpha,V)$. Since one does not have an explicit expression for $\IAA^{\alpha,V}_t(\omega)$ for $d>1$, we have to proceed differently for the general case: We will use the differential equation (\ref{pop1}) to rewrite $\mathscr{A}^{\alpha,V}_{t}(\omega)-\mathbf{1}$, and then use a uniform estimate for $\left\|\IAA^{\alpha,V}_t(\omega)\right\|$ (which is proved in lemma \ref{opo}) in order to derive an estimate that is similar to (\ref{dd2}).
\end{Remark}

{\it Proof of proposition \ref{hiu}.} We set $A:=A^{\alpha,V}$ and $\mathscr{A}:=\mathscr{A}^{\alpha,V}$. Since 
\begin{align}
\Id \mathscr{A}^{i}_{j} = (\mathscr{A} \underline{\Id} A)^{i}_{j}=\sum_{k} \mathscr{A}^{i}_{k} \underline{\Id} A^{k}_{j}=\sum_{k} \mathscr{A}^{i}_{k}\Id A^{k}_{j}+\sum_c\f{1}{2}\Id [\mathscr{A}^{i}_{k},A^{k}_{j}],\label{wwsx}
\end{align}
one has 
\begin{align}
\mathscr{A}^{i}_{j}-\delta^{i}_{j}=\sum_{k}\int \mathscr{A}^{i}_{k}\Id A^{k}_{j}+\f{1}{2}\sum_{k,l} \int \mathscr{A}^{i}_{l} \Id [A^{l}_{k},A^{k}_{j}].\label{aarr}
\end{align}
Furthermore, by the It\^{o} formula and $[X^{i}_t,X^{j}_t]=\delta^{i j} t$, $[X^{i}_t,t]=0$ for all $t>0$, one has
\begin{align}
A^{i}_{j}=\sum_{k} \int\alpha^{i}_{k,j}(X)\Id X^{k} + \f{1}{2}\int \sum_{k}\partial_{k}  \alpha^{i}_{k,j} (X) \Id t- \int V^{i}_{j}(X)\Id t 
\end{align}
and 
\begin{align}
 [A^{i}_{j},A^{k}_{l}]=   \sum_{m} \int \alpha^{i}_{m,j}(X)\alpha^{k}_{m,l}(X) \Id t,\label{wysx}
\end{align}
so that we arrive at
\begin{align}
\mathscr{A}^{i}_{j}-\delta^{i}_{j}
=&\sum_{k,l}\int \mathscr{A}^{i}_{k} \alpha^{k}_{l,j}(X)\Id X^{l} + \f{1}{2}\sum_{k,l}\int \mathscr{A}^{i}_{k}\partial_{l}  \alpha^{k}_{l,j} (X) \Id t\nn\\
&- \sum_{k}\int \mathscr{A}^{i}_{k}  V^{k}_{j}(X)\Id t    +\f{1}{2}\sum_{k,l,m} \int \mathscr{A}^{i}_{l} \alpha^{l}_{m,k}(X)\alpha^{k}_{m,j}(X) \Id t.
\end{align}
Let $t>0$. In order to use the It\^{o} isometry, we estimate the stochastic integrals by using Jensen's inequality as follows,
\begin{align}
\mathbb{E}^x \left[\left|\int^t_0 (\mathscr{A}_s)^{i}_{k} \alpha^{k}_{l,j}(X_s)\Id X^{l}_s\right|\right]^{2\f{1}{2}}&\leq \mathbb{E}^x\left[ \left| \int^t_0 (\mathscr{A}_s)^{i}_{k} \alpha^{k}_{l,j}(X_s)\Id X^{l}_s \right|^2\right]^{\f{1}{2}}\nn\\
&=\mathbb{E}^x\left[ \int^t_0\left|  (\mathscr{A}_s)^{i}_{k} \alpha^{k}_{l,j}(X_s)\right|^2\Id s\right]^{\f{1}{2}}.
\end{align}
By lemma \ref{opo}, there is a $C=C(d)>0$ such that for all $i,k=1,\dots,d$ and $s\geq 0$
\begin{align}
\left|(\mathscr{A}_s)^{i}_{k}\right|\leq C\>\>\text{ $\mathbb{P}^x$-a.s.,}\label{assp}
\end{align}
so that
\begin{align}
\mathbb{E}^x \left[\left| \mathscr{A}^{i}_{j}-\delta^{i}_{j}\right|\right]&\leq  C \sum_{k,l}\mathbb{E}^x \left[\int^t_0 \left| \alpha^{k}_{l,j}(X_s)\right|^2\Id s\right]^{\f{1}{2}}\nn\\
&+ \f{1}{2}C\sum_{k,l}\mathbb{E}^x \left[\int^t_0 \left| \partial_{l}  \alpha^{k}_{l,j} (X_s)\right| \Id s\right]\nn\\
&+ C  \sum_{k}\mathbb{E}^x \left[\int^t_0 \left|  V^{k}_{j}(X_s)\right|\Id s\right]  \nn\\
&  + \f{1}{2}C  \sum_{k,l,m} \mathbb{E}^x \left[\int^t_0 \left| \alpha^{l}_{m,k}(X_s)\alpha^{k}_{m,j}(X_s)\right| \Id s\right]
\end{align}
and the proof is complete by (\ref{ccr}). \vspace{0.5mm}

\hfill$\blacksquare$\vspace{2mm}

If one weakens the Kato assumption on $V$ in the previous proposition to a local Kato assumption, one still has:

\begin{Proposition}\label{aopq} Let $V$ be a nonnegative potential with
\[
V\in\mathcal{K}_{\mathrm{loc}}(\IR^n,\mathrm{Mat}(\IC^d))
\]
and let $\alpha\in\Omega^1(\IR^n, \mathscr{U}(d))$. Then for any compact $K\subset \IR^n$ one has
\begin{align}
\lim_{t\searrow 0}\sup_{x\in K}
\mathbb{E}^x\left[\left\|\mathscr{A}^{\alpha,V}_{t}-\mathbf{1}\right\|\right]=0.\label{eo}
\end{align}
\end{Proposition}

{\it Proof.} For any radius $r>0$ let $\zeta_{\mathrm{K}_r(0)}$ be the first exit time of $X$ from the open ball $\mathrm{K}_r(0)$. For any $t>0$ one has 
\begin{align}
&\sup_{x\in K}\mathbb{E}^x\left[\Big((1-1_{\{t<\zeta_{\mathrm{K}_r(0)}\}})+1_{\{t<\zeta_{\mathrm{K}_r(0)}\}}\Big)\left\|\mathscr{A}^{\alpha,V}_{t}-\mathbf{1}\right\|\right]\nn\\
&\leq 2\sup_{x\in K}\mathbb{E}^x\left[1-1_{\{t<\zeta_{\mathrm{K}_r(0)}\}}\right]+\sup_{x\in K}\mathbb{E}^x\left[1_{\{t<\zeta_{\mathrm{K}_r(0)}\}}\left\|\mathscr{A}^{\alpha,V}_{t}-\mathbf{1}\right\|\right],\label{qqpu}
\end{align}
where we have used lemma \ref{opo}. Since Levy's maximal inequality (as formulated in \cite{Si}) implies
\[
\sup_{x\in K}\mathbb{E}^x\left[1-1_{\{t<\zeta_{\mathrm{K}_r(0)}\}}\right]\to 0\>\>\text{ as $r\to \infty$ for any $t>0$,}
\]
taking $r\to\infty$ in (\ref{qqpu}) shows that it is sufficient to prove that for all $r>0$ one has
\begin{align}
\sup_{x\in \IR^n}
\mathbb{E}^x\left[1_{\{t<\zeta_{\mathrm{K}_r(0)}\}}\left\|\mathscr{A}^{\alpha,V}_{t}-\mathbf{1}\right\|\right]\to 0\>\>\text{ as $t\searrow 0$}.\label{rrc}
\end{align}
To this end, we first note that (\ref{wysx}) shows
\begin{align}
 (B^{\alpha,V})^i_j=\sum_k \int \alpha^i_{k,j}(X)\underline{\Id } X^k-\int V^i_j(X)\Id t+\f{1}{2}\sum_{k,l} \int \alpha^i_{l,k}(X)\alpha^k_{l,j}(X)\Id t.\label{ttv2}
\end{align}
We fix $t >0$, $r>0$ and let $\psi\in\mathrm{C}^{\infty}_0(\IR^n)$ be such that $\psi=1$ in $\mathrm{K}_r(0)$. It follows from (\ref{ttv2}) that $B^{\psi\alpha,\psi V}_s=B^{\alpha,V}_s$ in $ \{t<\zeta_{\mathrm{K}_r(0)}\}$ for all $0\leq s\leq t$. As a consequence, the expansion (\ref{for2}) for $\mathscr{A}^{\alpha,V}$ shows
\[
\mathbb{E}^x\left[1_{\{t<\zeta_{\mathrm{K}_r(0)}\}}\left\|\mathscr{A}^{\psi\alpha,\psi V}_{t}-\mathbf{1}\right\|\right]=\mathbb{E}^x\left[1_{\{t<\zeta_{\mathrm{K}_r(0)}\}}\left\|\mathscr{A}^{\alpha,V}_{t}-\mathbf{1}\right\|\right].
\] 
Since $\psi\alpha$ and $\psi V$ satisfy the assumptions of proposition \ref{hiu}, we have proved (\ref{rrc}). \vspace{0.5mm}

\hfill$\blacksquare$\vspace{2mm}

We now come to the main results of this paper. If $\alpha\in\Omega^1(\IR^n, \mathscr{U}(d))$, then the partial differential operator 
\begin{align}
 \tau(\alpha,0)\Psi= &-\f{1}{2}\Delta\Psi-\f{1}{2}  \sum^n_{j=1}\alpha_j^2\Psi-\f{1}{2}\sum^n_{j=1} (\partial_j\alpha_j) \Psi - \sum^n_{j=1}\alpha_j \partial_j\Psi,
\end{align}
defined initially for all $\Psi$ in the domain $\mathrm{D}(\tau(\alpha,0))=\mathrm{C}^{\infty}_0(\IR^n,\IC^d)$, is an essentially self-adjoint nonnegative \cite{Hess1} operator in the Hilbert space $\mathrm{L}^2(\IR^n,\IC^d)$ of (equivalence classes of) measurable functions $f=(f^1,\dots,f^d):\IR^n\to\IC^d$ such that
\[
\left\|f\right\|^2_{\mathrm{L}^2(\IR^n,\IC^d)}:=\int_{\IR^n} \left\|f(x)\right\|^2\Id x<\infty
\]
with scalar product
\[
\left\langle f,g\right\rangle_{\mathrm{L}^2(\IR^n,\IC^d)}= \int_{\IR^n} \left\langle f(x),g(x)\right\rangle \Id x,
\]
where $\left\langle  \bullet,\bullet \right\rangle $ denotes the Euclidean scalar product in $\IC^d$. We denote the quadratic form that corresponds to the closure $H(\alpha,0)\geq 0$ of $\tau(\alpha,0)$ with $q_{H(\alpha,0)}$. One has 
\begin{align}
&\mathrm{D}\left(  q_{\alpha,0} \right)=\left\{f\in \mathrm{L}^2(\IR^n,\IC^d) \left| \left(  \sum^n_{j=1}\left\|\partial_j f+\alpha_j f\right\|^2\right)^{\f{1}{2}} \in\mathrm{L}^2(\IR^n)\right\}\right.,\nn\\
&q_{\alpha,0}(f)=\f{1}{2}\int_{\IR^n}  \sum^n_{j=1}\left\|\partial_j f(x)+\alpha_j f(x)\right\|^2  \Id x,
\end{align}
which follows for example from proposition 8.13 in \cite{Br}, if one interprets $\Id+\alpha$ as a covariant derivative in $\IR^n\times\IC^d$. If $V$ is a nonnegative potential with
\[
V\in\mathcal{K}_{\mathrm{loc}}(\IR^n,\mathrm{Mat}(\IC^d)) \subset \mathrm{L}^1_{\mathrm{loc}}(\IR^n,\mathrm{Mat}(\IC^d)),
\]
then the KLMN-theorem (which we use in the sense of theorem 10.3.19 in \cite{Jo}) implies that the quadratic form given by 
\begin{align}
&\mathrm{D}(q_{\alpha,V}):=\mathrm{D}(q_{\alpha,0})\bigcap \left\lbrace f \left|  \int_{\IR^n} \left\langle V(x)f(x),f(x)\right\rangle  \Id x <\infty\right\rbrace\right.  \nn \\ 
&q_{\alpha,V}(f):=q_{\alpha,0}(f)+\int_{\IR^n} \left\langle V(x)f(x),f(x) \right\rangle  \Id x\nn
\end{align}
is densely defined, closed and nonnegative, and thus uniquely corresponds to a self-adjoint nonnegative operator $H(\alpha,V)$ in $\mathrm{L}^2(\IR^n,\IC^d)$. Differential operators of this type arise in nonrelativistic quantum mechanics, when one wants to describe the energy of Yang-Mills particles \cite{ho} \cite{stan}: These are particles with internal symmetries (modelled by a subgroup of $\mathrm{U}(d)$) that lead to a coupling with a matrix-valued Yang-Mills type field $\alpha$ as above. Also, the Zeeman-term in the Pauli operator \cite{lieb} (this operator models spin nonrelativistically) leads to the fact that the latter operator is of the form $H(\alpha,V)$, where here $\alpha$ is of the form $\tilde{\alpha}\otimes \mathbf{1}_{\mathrm{Mat}(\IC^d)}$, with some $\tilde{\alpha}\in\Omega(\IR^n,\mathscr{U}(1))$.

\vspace{1.2mm}

Note that under the above assumptions on $(\alpha,V)$, the expressions 
\[
\mathbb{E}^x\left[\mathscr{A}^{\alpha,V}_t f\left( X_t\right)  \right],\> x\in\IR^n,\> t>0,
\]
are well-defined (this follows from remark \ref{bem} and lemma \ref{opo}). As our first main result, we are going to prove the following Feynman-Kac type formula, which will be our main tool in the following:

\begin{Theorem}\label{aayv} Let $\alpha\in\Omega^1(\IR^n, \mathscr{U}(d))$ and let $V$ be a nonnegative potential with
\[
V\in\mathcal{K}_{\mathrm{loc}}(\IR^n,\mathrm{Mat}(\IC^d)).
\]
Then for any $t>0$, $f\in \mathrm{L}^2(\IR^n,\IC^d)$ and a.e. $x\in\IR^n$ one has 
\begin{align}
\mathrm{e}^{-t H(\alpha,V)}f(x)=\mathbb{E}^x\left[\mathscr{A}^{\alpha,V}_t f\left(X_t\right)  \right].\label{alpb}
\end{align}
\end{Theorem}

The proof of theorem \ref{aayv} will be given in section \ref{po}.\vspace{1.2mm}

As a first application of theorem \ref{aayv}, we are going to use proposition \ref{aopq} to prove the following theorem, which is our second main result:

\begin{Theorem}\label{wsx} Fix the assumptions of theorem \ref{aayv}. Then $\mathrm{e}^{-t H(\alpha,V)}f$ has a bounded continuous representative which is given by
\[
\IR^n\longrightarrow\IC^d,\>\>x\longmapsto \mathbb{E}^x\left[\mathscr{A}^{\alpha,V}_t f\left( X_t \right)  \right].
\]
In particular, any eigenfunction of $H(\alpha,V)$ can be chosen bounded and continuous.
\end{Theorem}

\begin{Remark}  If $n\leq 3$ and $V\in\mathrm{L}^2_{\mathrm{loc}}(\IR^n,\mathrm{Mat}(\IC^d))$, then one has 
\begin{align}
\mathrm{D}(H(\alpha,V))\subset \mathrm{H}^2_{\mathrm{loc}}(\IR^n,\IC^d),\label{zz}
\end{align}
(this follows for example from theorem 2.3 in \cite{Br}) which proves the continuity of the eigenfunctions in this case. In this sense, the continuity result from theorem \ref{wsx} extends this continuity to higher dimensions.
\end{Remark}

{\it Proof of theorem \ref{wsx}.} For any function $h:\IR^n\to \IC^d$ let
\[
 P^{\alpha,V}_th(x):=\mathbb{E}^x\left[\mathscr{A}^{\alpha,V}_t h(X_t) \right].
\]
If $f\in \mathrm{L}^2(\IR^n,\IC^d)$, then $P^{\alpha,V}_tf(x)$ is well-defined for all $t>0$, $x\in\IR^n$. Due to lemma \ref{opo}, the corresponding semigroup domination
\begin{align}
\left\|\mathbb{E}^x\left[\mathscr{A}^{\alpha,V}_{t} f(X_{t}) \right]\right\|\leq \mathbb{E}^x\left[\left\| f(X_{t })\right\| \right]\>\>\text{ for any $x\in\IR^n$,}
\end{align}
and the fact that $\mathbb{E}^{\bullet}\left[\left\| f(X_{t })\right\|\right]$ is bounded, we have that $ P^{\alpha,V}_tf$ is bounded for all $t>0$. \\                                                                                                  
In order to prove the asserted continuity, one can use the boundedness of $ P^{\alpha,V}_tf$ and the pointwise semigroup property of $(P^{\alpha,V}_t)_{t\geq 0}$ (which follows easily from (\ref{aaaq})), to see that we can assume that $f$ is bounded. Let us also note that for any $p\in [1,\infty]$ and $t>0$,
\[
P^{0,0}_t: \mathrm{L}^{p}(\IR^n,\IC^d)\longrightarrow  \mathrm{C}(\IR^n,\IC^d).
\]

Fix some arbitrary $t>0$ and let $s$ be such that $t\geq s>0$. By the above considerations, it is sufficient to prove that for any compact $K\subset \IR^n$ one has
\begin{align}
  \sup_{x\in K} \left\|\mathbb{E}^x\left[\tilde{f}(t-s,X_s)\right]-\mathbb{E}^x\left[\mathscr{A}^{\alpha,V}_t f(X_t) \right]\right\|\to 0\>\>\text{ as $s \searrow 0$},\label{aapi}
\end{align}
since
\[
\tilde{f}:[0,\infty)\times\IR^n\longrightarrow\IC^d,\>\>\tilde{f}(u,x):=\mathbb{E}^x\left[\mathscr{A}^{\alpha,V}_{u} f(X_{u}) \right]
\]
is bounded in $x$. We set $\mathscr{A}:=\mathscr{A}^{\alpha,V}$. Using the Markov property of the Brownian motion together with (\ref{aaaq}) shows that for any $x\in\IR^n$,
\begin{align}
\mathbb{E}^x\left[\tilde{f}(t-s,X_s)\right]-\mathbb{E}^x\left[\mathscr{A}_t f(X_t) \right]=\mathbb{E}^x \left[ \mathscr{A}_s^{-1} \mathscr{A}_t f(X_t)-\mathscr{A}_t
f(X_t)\right].\label{ecm}
\end{align}
Noting that by lemma \ref{ay11} one has
\begin{align}
\left\|\mathscr{A}_s^{-1} \mathscr{A}_t
\right\|\leq 1\>\>\text{ $\mathbb{P}^x$-a.s.,}\nn
\end{align}
we can estimate as follows,
\begin{align}
& \left\|\mathbb{E}^x \left[ \mathscr{A}_s^{-1} \mathscr{A}_t f(X_t)-\mathscr{A}_t
f(X_t)\right]\right\|\nn\\
&=\left\|\mathbb{E}^x \left[ 
(1-\mathscr{A}_{s})\mathscr{A}_s^{-1}\mathscr{A}_t f(X_t)\right]\right\|\leq \left\|f\right\|_{\infty}\mathbb{E}^x \left[ \left\|
(1-\mathscr{A}_{s})\right\| \right].\nn
\end{align}
Now (\ref{aapi}) follows from proposition \ref{aopq}.\vspace{0.5mm}

\hfill$\blacksquare$\vspace{2mm}

Our next aim will be to prove that $\mathrm{e}^{-t H(\alpha,V)}$ has a jointly continuous integral kernel. To this end, we need the Brownian bridge measure(s) $\mathbb{P}^{x,y}_t$: Let 
\[
p_t(x,y)=\f{1}{  (2\pi t)^{\f{n}{2}}}  \mathrm{e}^{-\f{\left\|x-y\right\|^2}{2t}}
\]
stand for the heat kernel of $\IR^n$. We fix arbitary $t>0$, $x,y\in \IR^n$ for the following considerations. Let $\Omega_t:=\mathrm{C}([0,t],\IR^n)$, let 
\begin{align}
 X^{(t)}:[0,t]\times \Omega_t\longrightarrow \IR^n \label{sy11}
\end{align}
be the canonical process and denote the corresponding $\sigma$-algebra and filtration with $\IFF^{(t)}$ and $(\IFF^{(t)}_s)_{0\leq s\leq t}$, respectively. The measure $\mathbb{P}^x_t$ stands for the Wiener measure on $(\Omega_t,\IFF^{(t)})$ which is concentrated on the paths $\omega:[0,t]\to\IR^n$ with $\omega(0)=x$. Then for any $x,y\in\IR^n$ the Brownian bridge measure $\mathbb{P}^{x,y}_t$ can be defined as the unique probability measure on $(\Omega_t,\IFF^{(t)})$ such that
\begin{align}
\left.\f{\Id \mathbb{P}^{x,y}_t}{\Id \mathbb{P}^x_t} \right|_{\IFF^{(t)}_s}= \f{p_{t-s}\left(X^{(t)}_s,y\right)}{p_t(x,y)}\>\>\text{ for any $s<t$}.\label{sqy}
\end{align}
The process (\ref{sy11}) is a well-defined continuous semi-martingale under $\mathbb{P}^{x,y}_t$, which is a Brownian bridge from $x$ to $y$ with terminal time $t$, so that $\mathbb{P}^{x,y}_t$ is concentrated on the set of paths $\omega:[0,t]\to\IR^n$ with $\omega(0)=x$ and $\omega(t)=y$. It is well-known (see for example corollary A.2 in \cite{Sz}) that the family $\mathbb{P}^{x,y}_t$ disintegrates $\mathbb{P}^x_t$ in the sense that
\begin{align}
\mathbb{P}^x_t(A) = \int_{\IR^n} \mathbb{P}^{x,y}_t(A)   p_t(x,y)\Id y\>\>\text{ for any $A\in \IFF^{(t)}$,}\label{bbq}
\end{align} 
and that for any $F\in\mathrm{L}^1(\mathbb{P}^{y,x}_t)$ one has the following time reversal property:
\[
\int_{\Omega_t}F(\omega(t-\bullet)) \mathbb{P}^{x,y}_t(\Id \omega) =\int_{\Omega_t}F(\omega) \mathbb{P}^{y,x}_t(\Id \omega).
\]

The local Kato class is compatible with the Brownian bridge measures in the following sense:

\begin{Remark}\label{rrl} If $V\in\mathcal{K}_{\mathrm{loc}}(\IR^n,\mathrm{Mat}(\IC^d))$, then by lemma C.8 in \cite{Bro1} one has
\begin{align}
\mathbb{P}^{x,y}_t\left\lbrace \int^t_0\left\|V\left(X^{(t)}_s \right)\right\| \Id s <\infty \right\rbrace=1.
\end{align}
\end{Remark}

The following definitions completely follow the construction of $\mathscr{A}^{\alpha,V}$: Let $\alpha\in\Omega^1(\IR^n,\mathscr{U}(d))$ and let $V\in\mathcal{K}_{\mathrm{loc}}(\IR^n,\mathrm{Mat}(\IC^d))$ be a potential. Remark \ref{rrl} and the fact that (\ref{sy11}) is a continuous semi-martingale under $\mathbb{P}^{x,y}_t$ show that 
\begin{align}
&A^{\alpha,V,(t)}:[0,t]\times \Omega_t\longrightarrow \mathrm{Mat}(\IC^d)\nn\\
&A^{\alpha,V,(t)}_s:=\sum^n_{j=1}\int^{s}_0 \alpha_j(X^{(t)}_r) \underline{\Id} X^{(t),j}_r-\int^{s}_0 V\left(X^{(t)}_r\right) \Id r\nn
\end{align}
is also a continuous semi-martingale under $\mathbb{P}^{x,y}_t$, so that the same is true for
\begin{align}
B^{\alpha,V,(t)}:[0,t]\times \Omega_t\longrightarrow \mathrm{Mat}(\IC^d),
\end{align}
which is defined in analogy to (\ref{z55}). If we furthermore set
\begin{align}
\mathscr{A}^{\alpha,V,(t)}_s &:= \mathbf{1}+\sum^{\infty}_{l=1} \int_{s\Delta_{l} } \Id
B^{\alpha,V,(t)}_{s_1}\cdots \Id B^{\alpha,V,(t)}_{s_{l}},
\end{align}
where the convergence is $\mathbb{P}^{x,y}_t$-a.s. uniformly in $[0,t]$, we have that 
\[
\mathscr{A}^{\alpha,V,(t)}:[0,t]\times\Omega_t\longrightarrow\mathrm{Mat}(\IC^d)
\]
is uniquely determined as the solution of
\begin{align}
 \mathscr{A}^{\alpha,V,(t)}_s = \mathbf{1}+\int^s_0 \mathscr{A}^{\alpha,V,(t)}_r \underline{\Id} A^{\alpha,V,(t)}_r\label{Pop1}
\end{align}
under $\mathbb{P}^{x,y}_t$. We will use the notation 
\[
\prod^{\longrightarrow}_{1\leq j\leq l} M_j:=M_1\cdots M_l\>\>\text{ for $M_1,\dots,M_l\in\mathrm{Mat}(\IC^d)$.}
\]

One has the following Hermitian symmetry:

\begin{Lemma}\label{edc} Let $\alpha\in\Omega^1(\IR^n,\mathscr{U}(d))$, let $V$ be a nonnegative potential with
\[
  V \in \mathcal{K}_{\mathrm{loc}}(\IR^n,\mathrm{Mat}(\IC^d)),
\]
and for any $t>0$ let  
\begin{align}
&\mathrm{e}^{-t H(\alpha,V)}(\bullet,\bullet):\IR^n\times\IR^n\longrightarrow\mathrm{Mat}(\IC^d),\nn\\
&\mathrm{e}^{-t H(\alpha,V)}(x,y):=  \f{1}{  (2\pi t)^{\f{n}{2}}}  \mathrm{e}^{-\f{\left\|x-y\right\|^2}{2t}} \mathbb{E}^{x,y}_t\left[\mathscr{A}^{\alpha,V,(t)}_t \right].
\end{align}
Then $\mathrm{e}^{-t H(\alpha,V)}(x,y)$ is well-defined for all $t>0$, $x,y\in \IR^n$ and one has 
\begin{align}
\mathrm{e}^{- t H(\alpha,V)}(y,x)=\mathrm{e}^{- t H(\alpha,V)}(x,y)^*.\label{gb11}
\end{align}
\end{Lemma}

\begin{Remark} Let $d=1$. Using that $\mathbb{P}^{x,y}_t$ is equivalent to $\mathbb{P}^x_t$ on $\IFF^{(t)}_s$ for all $0\leq s<t$, it follows (from taking $s\nearrow t$ and from the fact that $X^{(t)}$ is a continuous semi-martingale under $\mathbb{P}^{x,y}_t$) that for all $j,k=1,\dots,n$ one has
\[
[X^{(t),j},X^{(t),k}]_s=\delta^{jk} s\>\>\text{ $\mathbb{P}^{x,y}_t$-a.s. for all $0\leq s\leq t$.}
\]
As a consequence, the It\^{o}-formula gives 
\begin{align}
& \mathscr{A}^{\alpha,V,(t)}_s\nn\\
&=\exp\left( \mathrm{i}\sum^n_{j=1}\int^{s}_0 \tilde{\alpha}_j(X^{(t)}_r) \Id X^{(t),j}_r+\f{\mathrm{i}}{2}\int^s_0\mathrm{div}(\tilde{\alpha})(X^{(t)}_r)\Id r-\int^{s}_0 V\left(X^{(t)}_r\right) \Id r\right)\label{zqay}
\end{align}
$\mathbb{P}^{x,y}_t$-a.s. for all $0\leq s\leq t$. In particular, (\ref{gb11}) becomes a simple consequence of the time reversal property of the Brownian bridge in this case. For the general case, we will use a result \cite{Em2} by Emery, which states that $\IAA^{\alpha,V,(t)}$ can be approximated by stochastic product integrals.
\end{Remark}

{\it Proof of lemma \ref{edc}.} The well-definedness of $\mathrm{e}^{-t H(\alpha,V)}(x,y)$ follows from remark \ref{rrl} and lemma \ref{opo}:
\[
\left\|\mathrm{e}^{-t H(\alpha,V)}(x,y)\right\|\leq  \f{1}{  (2\pi t)^{\f{n}{2}}}.
\]

We set $\mathscr{A}^{(t)}:=\mathscr{A}^{\alpha,V,(t)}$ and $B^{(t)}:=B^{\alpha,V,(t)}$. The time reversal property of the Brownian bridge measure implies
\[
 \int_{\Omega_t} \mathscr{A}^{(t)}_t(\omega) \ \mathbb{P}^{y,x}_t(\Id \omega)=  \int_{\Omega_t} \mathscr{A}^{(t)}_t(\omega(t-\bullet)) \ \mathbb{P}^{x,y}_t( \Id\omega),
\]
so that it is sufficient to prove 
\begin{align}
 \mathscr{A}^{(t),*}_t(\omega(t-\bullet))=\mathscr{A}^{(t)}_t(\omega)\>\>\text{ for $\mathbb{P}^{y,x}_t$-a.e. $\omega\in\Omega_t$.}\label{uiu}
\end{align} 
We can proceed as follows in order to prove the latter equality: For any partition 
\[
\sigma=\Big\{0=t_0<t_1<t_2\dots<t_m=t\Big\}
\]
of $[0,t]$ we define 
\begin{align}
\mathscr{A}^{(t),\sigma}_t:=\left( \mathbf{1}+B^{(t)}_{t_0}\right) \prod^{\longrightarrow}_{1\leq j\leq m} \left( \mathbf{1}+B^{(t)}_{t_j}-B^{(t)}_{t_{j-1}}\right). \label{pqi}
\end{align}
Analogously to (\ref{ttv2}) one has
\begin{align}
 (B^{(t)})^i_j=&\sum_l \int \alpha^i_{l,j}(X^{(t)})\underline{\Id } X^{(t),l}-\int V^i_j(X^{(t)})\Id s\nn\\
&+\f{1}{2}\sum_{k,l} \int \alpha^i_{l,k}(X^{(t)})\alpha^k_{l,j}(X^{(t)})\Id s. \label{t1v2}
\end{align}

By \cite{Em2}, p.256, the family of random variables $(\mathscr{A}^{(t),\sigma}_t)_{\sigma}$ converges in probability (with respect to $\mathbb{P}^{y,x}_t$) to $\mathscr{A}^{(t)}_t$ as $|\sigma|\to 0$. Now the key observation for proving (\ref{uiu}) is the following: Since $\alpha^*_j=-\alpha_j$, $j=1,\dots,n$, and $V=V^*$, approximating the integrals in (\ref{t1v2}) with Riemann-type sums implies
\begin{align}
\ B^{(t),*}_s(\omega(t-\bullet))=B^{(t)}_t(\omega)-B^{(t)}_{t-s}(\omega)\>\>\text{ for $\mathbb{P}^{y,x}_t$-a.e. $\omega\in\Omega_t$, $0\leq s\leq t$}.\label{tztz}
\end{align}
Now (\ref{uiu}) follows from (\ref{tztz}) and the adjoint version of formula (\ref{pqi}). \vspace{0.5mm}

\hfill$\blacksquare$\vspace{2mm}

Being equipped with this result, we can use proposition \ref{aopq} to prove our third main result:

\begin{Theorem}\label{rfv} Fix the assumptions of theorem \ref{aayv}. \vspace{1.2mm}

{\rm a)} The map $\mathrm{e}^{-t H(\alpha,V)}(\bullet,\bullet)$ represents an integral kernel of $\mathrm{e}^{-t H(\alpha,V)}$ in the sense that for all $f\in\mathrm{L}^2(\IR^n,\IC^d)$ and a.e. $x\in\IR^n$ one has
\begin{align}
\mathrm{e}^{-t H(\alpha,V)}f(x)=\int_{\IR^n} \mathrm{e}^{-t H(\alpha,V)}(x,y) f(y) \Id y.
\end{align}

{\rm b)} The map 
\[
(0,\infty)\times \IR^n\times\IR^n\longrightarrow\mathrm{Mat}(\IC^d),\>(t,x,y)\longmapsto \mathrm{e}^{- t H(\alpha,V)}(x,y)
\]
is bounded in $(x,y)$ and jointly continuous in $(t,x,y)$.\vspace{1.2mm}

{\rm c)} It holds that
\begin{align}
\mathrm{tr}_{\mathrm{L}^2(\IR^n,\IC^d)}\left(\mathrm{e}^{- t H(\alpha,V)}\right)=\int_{\IR^n}\mathrm{tr}_{\mathrm{Mat}(\IC^d)}\left(\mathrm{e}^{- t H(\alpha,V)}(x,x)\right) \Id x,\label{exm}
\end{align}
as a number in $[0,\infty]$.
\end{Theorem}

{\it Proof.} a) Let 
\[
 \Pi_t:\Omega\longrightarrow \Omega_t,\>\> \Pi_t(\omega)=\omega\mid_{[0,t]}
\]
denote the canonical projection. Since $X^{(t)}$ is a continuous semi-martingale under $\mathbb{P}^x_t$ (in fact, a Brownian motion starting in $x$), the expansion for $\mathscr{A}^{\alpha,V,(t)}$ converges with respect to $\mathbb{P}^x_t$ and one has 
\begin{align}
\mathscr{A}^{\alpha,V}_s=\mathscr{A}^{\alpha,V,(t)}_s\circ \Pi_t\>\>\text{ $\mathbb{P}^x$-a.s. for all $0\leq s\leq t$.}\label{zb1}
\end{align}
It follows from (\ref{zb1}), (\ref{bbq}) and $X^{(t)}_t=y$ $\mathbb{P}^{x,y}_t$-a.s. that
\begin{align}
\mathbb{E}^x\left[  \mathscr{A}^{\alpha,V}_t f(X_t)\right]&= \mathbb{E}^x\left[ \left( \mathscr{A}^{\alpha,V,(t)}_t f\left(X^{(t)}_t\right) \right)   \circ \Pi_t\right]\nn\\
& =\mathbb{E}^x_t\left[ \mathscr{A}^{\alpha,V,(t)}_t f\left( X^{(t)}_t\right) \right]\nn\\
&=\int_{\IR^n} p_t(x,y) \mathbb{E}^{x,y}_t\left[ \mathscr{A}^{\alpha,V,(t)}_t \right]  f(y) \Id y\nn. 
\end{align}

b) We set $\mathscr{A}^{(t)}:=\mathscr{A}^{\alpha,V,(t)}$ and $A^{(t)}:=A^{\alpha,V,(t)}$ for any $t>0$. The asserted boundedness has already been checked in the proof of lemma \ref{edc}. For the continuity, let $K\subset \IR^n$ be an arbitrary compact subset, and let $\tau_1\leq\tau_2$ be arbitrary positive real numbers. In view of lemma \ref{edc}, one can go through the same steps as in the proof of theorem 6.1 in \cite{Bro1} to see that it is sufficient to prove that
\begin{align}
\lim_{s\searrow  0} \>\sup_{\tau_1\leq t\leq \tau_2}\> \sup_{x,y\in K}  \left\|\Psi(t,s,x,y)\right\|=0, \label{ekl}
\end{align}
and that for all $0<s <  \tau_1$,
\begin{align}
\lim_{r\searrow 0}\>\> \sup_{\tau_1\leq t\leq \tilde{t}\leq \tau_2,\>|t-\tilde{t}|<r}\> \>\>\sup_{x,y\in K,\>\left\|y-\tilde{y}\right\|<r}  \left\|\Phi(t,\tilde{t},s,x,y,\tilde{y})\right\|=0,\label{ekl2}
\end{align}
where 
\begin{align}
&\Psi:[\tau_1,\tau_2]\times (0,\tau_1)\times K\times K \longrightarrow \mathrm{Mat}(\IC^d)\nn\\
&\Psi(t,s,x,y):=p_t(x,y)\mathbb{E}^{x,y}_t\left[\mathscr{A}^{(t)}_t-\mathscr{A}^{(t)}_{t-s} \right],\nn\\
&\Phi:[\tau_1,\tau_2]\times[\tau_1,\tau_2]\times  (0,\tau_1)\times K\times K\times K  \longrightarrow \mathrm{Mat}(\IC^d)\nn\\
&\Phi(t,\tilde{t},s,x,y,\tilde{y}):=p_{\tilde{t}}(x,\tilde{y})\mathbb{E}^{x,\tilde{y}}_{\tilde{t}}\left[\mathscr{A}^{(\tilde{t})}_{(t-s)\theta(\tilde{t}-t+s)}\right]-p_t(x,y)\mathbb{E}^{x,y}_t\left[\mathscr{A}^{(t)}_{t-s} \right],\nn
\end{align}
and where $\theta:\IR\to [0,\infty)$ stands for the Heaviside function.\vspace{1.2mm}

{\it Proof of (\ref{ekl}):} One has 
\begin{align}
\left\| \Psi(t,s,x,y)\right\|&\leq p_t(x,y) \mathbb{E}^{x,y}_t\left[ \left\|\mathscr{A}^{(t)}_t-\mathscr{A}^{(t)}_{t-s} \right\|\right]\nn\\
&= p_t(x,y) \mathbb{E}^{x,y}_t\left[ \left\|\mathscr{A}^{(t)}_{t-s}\left(  \mathscr{A}^{(t),-1}_{t-s}\mathscr{A}^{(t)}_t-\mathbf{1}\right)  \right\|\right]\nn\\
&\leq   \ p_t(x,y) \mathbb{E}^{x,y}_t\left[\left\|    \mathscr{A}^{(t),-1}_{t-s}\mathscr{A}^{(t)}_t-\mathbf{1}  \right\|\right],
\end{align}
where we have used that
\[
 \left\|\mathscr{A}^{(t)}_{t-s}\right\|\leq 1 \>\>\text{$\mathbb{P}^{x,y}_t$-a.s.}
\]
by lemma \ref{opo}. The time reversal property of the Brownian bridge measure shows 
\begin{align}
& \mathbb{E}^{x,y}_t\left[ \left\|    \mathscr{A}^{(t),-1}_{t-s}\mathscr{A}^{(t)}_t-\mathbf{1}   \right\|\right]\nn\\
&=\int_{\Omega_t} \left\|   \mathscr{A}^{(t),-1}_{t-s}(\omega(t-\bullet))\mathscr{A}^{(t)}_t(\omega(t-\bullet))-\mathbf{1} \right\| \mathbb{P}^{y,x}_t(\Id \omega).
\end{align}

Using the identity
\begin{align}
  \mathscr{A}^{(t),-1}_{t-s}(\omega(t-\bullet))\mathscr{A}^{(t)}_t(\omega(t-\bullet))= \mathscr{A}_s^{(t),*}(\omega)\>\>\text{ for $\mathbb{P}^{y,x}_t$-a.e. $\omega\in\Omega_t$},\label{wpo}
\end{align}
which we are going to prove in a moment, and using (\ref{sqy}) and (\ref{zb1}) we arrive at 
\begin{align}
 \left\|\Psi(t,s,x,y)\right\|&\leq\>  p_t(x,y) \mathbb{E}^{y,x}_t\left[\left\|     \mathscr{A}^{(t),*}_{s}-\mathbf{1}  \right\|\right]\nn\\
&=\>    (2\pi (t-s))^{-\f{n}{2}}   \mathbb{E}^{y}\left[\mathrm{e}^{-\f{\left\|y-X_s\right\|^2}{2(t-s)}}\left\|     \mathscr{A}^{*}_{s}-\mathbf{1}  \right\|\right]\nn\\
&\leq \> (2\pi (t-s))^{-\f{n}{2}}  \mathbb{E}^{y}\left[\left\|     \mathscr{A}^{*}_{s}-\mathbf{1}  \right\|\right].
\end{align}
Now (\ref{ekl}) is implied by proposition \ref{aopq}.\\
It remains to prove (\ref{wpo}): Note that if $d=1$, then this formula follows directly from (\ref{zqay}) and $\mathrm{e}^{z_1+z_2}=\mathrm{e}^{z_1}\mathrm{e}^{z_2}$. For the general case, we will (analogously to the proof of lemma \ref{tcy}) use the following trick: We will prove that for fixed $t$, both sides of (\ref{wpo}) solve the same initial value problem with respect to $s$. To this end, fix some arbitrary $t>0$, $x,y\in\IR^n$, and let the process 
\[
\tilde{\mathscr{A}}^{(t)}:[0,t]\times \Omega_t\longrightarrow \mathrm{Mat}(\IC^d)
\]
be given by $\tilde{\mathscr{A}}^{(t)}_s(\omega)=\mathscr{A}^{(t)}_{t-s}(\omega(t-\bullet))$. Then, with respect to $\mathbb{P}^{y,x}_t$, one has 
\begin{align}
 \tilde{\mathscr{A}}^{(t)}_s(\omega)=\mathbf{1}+\left( \int^{t-s}_0 \mathscr{A}^{(t)}_r \underline{\Id }A^{(t)}_r \right) (\omega(t-\bullet)).\label{zgb}
\end{align}
As in (\ref{tztz}) one sees
\begin{align}
A^{(t),*}_s(\omega(t-\bullet))=A^{(t)}_t(\omega)-A^{(t)}_{t-s}(\omega)\>\>\text{ for $\mathbb{P}^{y,x}_t$-a.e. $\omega\in\Omega_t$,}\label{zv1}
\end{align}
so using the adjoint version of (\ref{zv1}) and approximating the Stratonovic integral in (\ref{zgb}) with Riemann sums easily implies the first identity in
\begin{align}
\left( \int^{t-s}_0 \mathscr{A}^{(t)} \underline{\Id }A^{(t)} \right) (\omega(t-\bullet))&=\left( \int^{t}_s \tilde{\mathscr{A}}^{(t)}_r \underline{\Id }A^{(t),*}_r \right) (\omega)\nn\\
&= \left( \int^{t}_0 \tilde{\mathscr{A}}^{(t)}_r \underline{\Id }A^{(t),*}_r \right) (\omega)-\left( \int^{s}_0 \tilde{\mathscr{A}}^{(t)}_r \underline{\Id }A^{(t),*}_r \right) (\omega).\nn
\end{align}
Thus, $\tilde{\mathscr{A}}^{(t)}$ is uniquely determined as the solution of $\Id \tilde{\mathscr{A}}^{(t)}_s = -\tilde{\mathscr{A}}^{(t)}_s \underline{\Id }A^{(t),*}_s$ with initial value $\tilde{\mathscr{A}}^{(t)}_0(\omega)= \mathscr{A}^{(t)}_t(\omega(t-\bullet))$, which shows 
\[
\tilde{\mathscr{A}}^{(t)}_s(\omega)=\mathscr{A}^{(t)}_t(\omega(t-\bullet))\mathscr{A}^{(t),*,-1}_s(\omega)\>\>\text{ for $\mathbb{P}^{y,x}_t$-a.e. $\omega \in\Omega_t$}
\]
and (\ref{wpo}) is proved.\vspace{1.2mm}

{\it Proof of (\ref{ekl2}):} In view of (\ref{ekl2}) let $t\leq \tilde{t}$. Using (\ref{sqy}) and (\ref{zb1}) we have
\begin{align}
&\Phi(t,\tilde{t},s,x,y,\tilde{y})\nn\\ &=\mathbb{E}^x\left[ (2\pi)^{-\f{n}{2}} \left(   (\tilde{t}-t+s)^{-\f{n}{2}}\mathrm{e}^{-\f{\left\|X_{t-s}-\tilde{y}\right\|^2}{2(\tilde{t}-t+s)}} -s^{-\f{n}{2}}\mathrm{e}^{-\f{\left\|X_{t-s}-y\right\|^2}{2 s}}\right)  \mathscr{A}_{t-s}\right],
\end{align}
so that Jensen's inequality gives
\begin{align}
&\left\|\Phi(t,\tilde{t},s,x,y,\tilde{y})\right\|^2\nn\\
&\leq  (2\pi)^{-n}  \mathbb{E}^x\left[  \left(   (\tilde{t}-t+s)^{-\f{n}{2}}\mathrm{e}^{-\f{\left\|X_{t-s}-\tilde{y}\right\|^2}{2(\tilde{t}-t+s)}} -s^{-\f{n}{2}}\mathrm{e}^{-\f{\left\|X_{t-s}-y\right\|^2}{2 s}}\right)^2 \right].
\end{align}
Now the proof of theorem 6.1 in \cite{Bro1} can be copied word by word. \vspace{1.2mm}

c) This formula follows directly from the continuity of the integral kernel and well-known algebraic arguments (see for example the proof proposition 12 in \cite{weis}). \vspace{0.5mm}

\hfill$\blacksquare$\vspace{2mm}

\textbf{Acknowledgements.} The research has been financially supported by the Bonner Internationale Graduiertenschule.

\section{ Proof of theorem \ref{aayv}.}\label{po}

For any potential $W:\IR^n\to\mathrm{Mat}(\IC^d)$ which satisfies (\ref{bb11}) (with $V$ replaced with $W$) for all $x\in \IR^n$, we define the process 
\[
\tilde{\mathscr{A}}^{\alpha,W}:[0,\infty)\times\Omega\longrightarrow \mathrm{Mat}(\IC^d)
\]
as the path ordered exponential 
\[
\tilde{\mathscr{A}}^{\alpha,W}_t=\mathbf{1}+\sum^{\infty}_{l=1}\int_{t\Delta_l}\prod^{\longrightarrow}_{1\leq j\leq l} \left( -\mathscr{A}^{\alpha,0}_{t_j} W(X_{t_j})\mathscr{A}^{\alpha,0,-1}_{t_j} \right)  \Id t_1\dots \Id t_l.
\] 
Then $\tilde{\mathscr{A}}^{\alpha,W}$ is nothing but the pathwise weak solution \cite{dollard} of
\begin{align}
\f{\Id}{\Id t} \tilde{\mathscr{A}}^{\alpha,W}_t =-\tilde{\mathscr{A}}^{\alpha,W}_t  \mathscr{A}^{\alpha,0}_t W(X_t)\mathscr{A}^{\alpha,0,-1}_t,\>\>\tilde{\mathscr{A}}^{\alpha,W}_0=\mathbf{1},
\end{align}
and the Stratonovic product rule implies the following formula for any $x\in\IR^n$, 
\begin{align}
\mathscr{A}^{\alpha,W}_t=\tilde{\mathscr{A}}^{\alpha,W}_t\mathscr{A}^{\alpha,0}_t\>\>\text{ $\mathbb{P}^x$-a.s.} \label{deco}
\end{align}
Furthermore, the unitarity $\mathscr{A}^{\alpha,0,-1}=\mathscr{A}^{\alpha,0,*}$ (by lemma \ref{opo} a)) combined with Gronwall's lemma implies the following inequality for any $x\in\IR^n$,
\begin{align}
\left\|\tilde{\mathscr{A}}^{\alpha,W}_t\right\|\leq \mathrm{e}^{\int^t_0 \left\|W(X_s)\right\|\Id s} \>\>\text{ $\mathbb{P}^x$-a.s.}\label{bb33}
\end{align}

We fix arbitrary $t>0$ and $f\in \mathrm{L}^{2}(\IR^n,\IC^d)$. The remaining proof can be divided into three steps, and is modelled after the proof of theorem 1.3 in \cite{Gue}.\vspace{1.3mm}

{\it Step 1. Assume that $V$ is a potential in $\mathrm{C}_{b}(\IR^n,\mathrm{Mat}(\IC^d))$, the space of continuous bounded functions $\IR^n\to\mathrm{Mat}(\IC^d)$. }\vspace{1mm}

The operator
\[
P^{\alpha,V}_t:\mathrm{L}^2(\IR^n,\IC^d)\longrightarrow \mathrm{L}^2(\IR^n,\IC^d),\>\>P^{\alpha,V}_t h(x):=\mathbb{E}^x\left[ \IAA^{\alpha,V}_t h(X_t) \right]
\]
is a well-defined bounded linear operator in $\mathrm{L}^2(\IR^n,\IC^d)$. If $\psi\in\mathrm{C}^{\infty}_0(\IR^n,\IC^d)$, then a straightforward calculation, which uses the It\^{o} formula repeatedly, shows that for any $x\in\IR^n$, one has the following equality $\mathbb{P}^x$-a.s.,
\begin{align}
&\IAA^{\alpha,V}_t \psi(X_t)=\text{ [a martingale which starts from $0$]} + \psi(x)+ \nn\\
&\int^t_0 \IAA^{\alpha,V}_s \Delta\psi(X_s)\Id s+\int^t_0 \IAA^{\alpha,V}_s\sum^n_{j=1} (\partial_j\alpha_j(X_s)) \psi(X_s)\Id s\nn\\
&+2\int^t_0 \IAA^{\alpha,V}_s \sum^n_{j=1} \alpha_j(X_s) \partial_j \psi(X_s)\Id s +\int^t_0 \sum^n_{j=1}\alpha^2_j(X_s) \psi(X_s)\Id s
\nn\\
&- \int^t_0 \IAA^{\alpha,V}_s V(X_s)\Id s,\nn
\end{align} 
so that taking $\mathbb{E}^x[\bullet]$ in this equation implies
\begin{align}
P^{\alpha,V}_t \psi(x)=\psi(x) -\int^t_0 P^{\alpha,V}_s H(\alpha,V)\psi(x) \Id s.
\end{align}
This shows $P^{\alpha,V}_t\psi=\mathrm{e}^{-t H(\alpha,V)}\psi$ so that the boundedness of $P^{\alpha,V}_t$ implies $P^{\alpha,V}_tf=\mathrm{e}^{-t H(\alpha,V)}f$, the Feynman-Kac formula. \vspace{1.3mm}

{\it Step 2. Assume that $V$ is a potential in $ \mathrm{L}^{\infty}(\IR^n,\mathrm{Mat}(\IC^d))$. }\vspace{1mm}

Using Friedrichs mollifiers as in \cite{Jo}, p.280, one finds a sequence $(V_m)$ of continuous bounded potentials with 
\begin{align}
V_m(x)\to V(x)\text{ as $m\to\infty$,}\>\>\left\|V_m(x)\right\| \leq C(d)\left\|V\right\|_{\infty}\>\>\text{ for a.e. $x\in\IR^n$}.\label{wx00}
\end{align}
It follows from (\ref{wx00}) and dominated convergence that for any $\psi\in \mathrm{C}^{\infty}_0(\IR^n,\IC^d)$,
\begin{align}
\left\|H(\alpha,V_m)\psi- H(\alpha,V)\psi\right\|_{\mathrm{L}^2(\IR^n,\IC^d)}\to 0\>\>\text{ as $m\to\infty$.}
\end{align}
As a consequence, theorem VIII 25 and theorem VIII 20 from \cite{Re} show that we may assume
\begin{align}
 \mathrm{e}^{-tH(\alpha,V_m)}f(x)\to \mathrm{e}^{-tH(\alpha,V)}f(x)\>\>\text{ as $m\to\infty$ for a.e. $x\in\IR^n$.}\label{ss99}
\end{align}
On the other hand, the decomposition (\ref{deco}) combined with lemma \ref{ay11} b) implies (keeping in mind that $\mathscr{A}^{\alpha,0}$ is unitary) 
\begin{align}
  \left\|\IAA^{\alpha,V_m}_t -\IAA^{\alpha,V}_t \right\| &\leq \left\|\tilde{\IAA}^{\alpha,V_m}_t -\tilde{\IAA}^{\alpha,V}_t \right\|\nn\\
&\leq  \mathrm{e}^{2\int^t_0 \left\|V_m(X_s )\right\|\Id s+\int^t_0 \left\|V(X_s)\right\|\Id s }\int^t_0  \left\|V_m(X_s)-   V(X_s)\right\|\Id s,
\end{align}
so that by (\ref{wx00}) and dominated convergence, 
\begin{align}
&  \left\|\IAA^{\alpha,V_m}_t f(X_t)-\IAA^{\alpha,V}_t f(X_t)\right\|\nn\\
&\leq \left\|f(X_t)\right\| \mathrm{e}^{ 2(C(d)+1)\int^t_0 \left\|V(X_s)\right\|\Id s }\int^t_0  \left\|V_m(X_s)-   V(X_s)\right\|\Id s\nn\\
&\to 0\>\>\text{ as $m\to\infty$, $\mathbb{P}^x$-a.s. for any $x\in\IR^n$.}\label{b5nd}
\end{align}
Furthermore, (\ref{bb33}) and (\ref{wx00}) imply \footnote{Note that $\mathbb{E}^x\left[\left\|f(X_t)\right\| \right]=\mathrm{e}^{t\Delta}\left\|f(\bullet)\right\|(x)<\infty$.}
\[
\left\|\IAA^{\alpha,V_m}_t f(X_t)\right\|\leq \mathrm{e}^{ C(d)\int^t_0 \left\|V(X_s)\right\|\Id s }\left\|f(X_t)\right\|\leq \mathrm{e}^{ C(d)t\left\|V\right\|_{\infty} }\left\|f(X_t)\right\|\in\mathrm{L}^1(\mathbb{P}^x) 
\]
so that by (\ref{b5nd}) we may use dominated convergence to deduce
\begin{align}
\mathbb{E}^x\left[ \IAA^{\alpha,V_m}_t f(X_t)\right]\to \mathbb{E}^x\left[ \IAA^{\alpha,V}_t f(X_t)\right] \>\>\text{ as $m\to\infty$ for any $x\in\IR^n$,}\label{ggbv2}
\end{align}
and the Feynman-Kac formula for essentially bounded potentials follows from combining (\ref{ss99}) with the result from step 1. \vspace{1.3mm}

{\it Step 3. Assume that $V$ is potential with $0\leq V\in \mathcal{K}^{\mathrm{loc}}(\IR^n,\mathrm{Mat}(\IC^d))$. }\vspace{1mm}

Let $U:\IR^n\to \mathrm{U}(d)$ be a measurable function with 
\[
V(x)=U^*(x)\mathrm{diag}(v_1(x),\dots,v_d(x))U(x) \>\>\text{ for a.e. $x\in\IR^n$,}
\]
and for any $m\in\IN$ we define an essentially bounded nonnegative potential $V_m:\IR^n\to\mathrm{Mat}(\IC^d)$ by setting
\[
V_m(x):= U^*(x)\mathrm{diag}\left( v^{(m)}_1(x),\dots,v^{(m)}_d(x)\right) U(x),
\]
where $v^{(m)}_j(x):= \min\{v_j(x),m\}$. Note that we again have (\ref{wx00}) and that by monotone convergence of quadratic forms we may also assume (\ref{ss99}) (see \cite{Re}, theorem S.14 on p.373). On the other hand, (\ref{wx00}) shows that one can use the same arguments as in the proof of step 2 to deduce (\ref{b5nd}). Furthermore, since $V_m\geq 0$, it follows from lemma \ref{opo} a) that 
\[
 \left\|\IAA^{\alpha,V_m}_t f(X_t)\right\|\leq \left\|f(X_t)\right\|\in\mathrm{L}^1(\mathbb{P}^x),
\]
so that we also have (\ref{ggbv2}). Now the general Feynman-Kac formula follows from (\ref{ss99}) and step 2. \vspace{0.5mm}

\hfill$\blacksquare$\vspace{2mm}

\section{Appendix A}

We prove two auxiliary results here. \vspace{1.2mm}

The first assertion gives estimates on the solutions of certain matrix-valued ordinary linear differential equations: Fix $t_0\geq 0$ and let 
\[
F\in \mathrm{L}^1_{\mathrm{loc}}([t_0,\infty),\mathrm{Mat}(\IC^d)). 
\]
Then a standard use of the Banach fixed point theorem shows that there is a unique weak (= absolutely continuous) solution $Y:[t_0,\infty)\to  \mathrm{Mat}(\IC^d)$ of the ordinary initial value problem
\[
\f{\Id}{\Id s} Y(s)=Y(s)F(s),\>\>Y(t_0)=\mathbf{1}.
\]

We will write $ \left\langle  \bullet,\bullet \right\rangle$ for the Euclidean inner product in $\IC^d$ and $\left\|\bullet\right\|$ will stand for the induced norm on $\IC^d$ and also for the induced operator norm on $\mathrm{Mat}(\IC^d)$.

\begin{Lemma}\label{ay11} {\rm a)} Assume that $F(s)$ is Hermitian and that there exists a real-valued function $c\in\mathrm{L}^1_{\mathrm{loc}}[t_0,\infty)$ such that $F(s)\leq c(s)$ for a.e. $s\geq t_0$. 
Then 
\[
\left\|Y(t)\right\|\leq  \mathrm{e}^{\int^t_{t_0} c(r) \Id r}\>\>\text{ for any $t\geq t_0$}.
\]
{\rm b)} Let $F_1$, $F_2\in \mathrm{L}^1_{\mathrm{loc}}([t_0,\infty),\mathrm{Mat}(\IC^d))$ and let 
\[
Y_1, Y_2:[t_0,\infty)\longrightarrow  \mathrm{Mat}(\IC^d)
\]
be the unique solutions of the ordinary initial value problems
\[
\f{\Id}{\Id s} Y_j(s)= Y_j(s)F_j(s),\>\>Y_j(t_0)=\mathbf{1}\>\>\text{ for }j=1,2.
\]
The following inequality holds for all $t\geq t_0$,
\begin{align}
\left\|Y_1(t)-Y_2(t)\right\|\leq   \> & \mathrm{e}^{2\int^t_{t_0} \left\|F_{1}(s )\right\|\Id s+\int^t_{t_0} \left\|F_{2}(s)\right\|\Id s }\int^t_{t_0}  \left\|F_{1}(s)-   F_{2}(s)\right\|\Id s.\nn
\end{align}
\end{Lemma}

{\it Proof.} The lemma is included in proposition B.1 and proposition B.2 of \cite{Gue}. We give the short proof for the convinience of the reader.\vspace{1.2mm}

a) Let $e_1,\dots,e_k$ be the standard orthonormal basis of $\IC^d$. Since $\left\|Y^*\right\|=\left\|Y\right\|$, we can assume that 
\[
\f{\Id}{\Id s} Y(s) f_j =F(s)Y(s)f_j,\>\> Y(t_0)=\mathbf{1},
\]
so
\begin{align}
\f{\Id }{\Id s} \left\|Y(s)f_j\right\|^2 &= 2  \left\langle  F(s) (Y(s)f_j), Y(s)f_j \right\rangle\nn\\
&\leq 2  c(s) \left\|Y(s)f_j\right\|^2\>\>\text{ for a.e. $s\geq t_0$},
\end{align}
and the assertion follows from the Gronwall lemma.\vspace{1.2mm}

b) $Y_1(s)$ and $Y_2(s)$ are invertible for any $s\geq t_0$ and
\[
\f{\Id}{\Id s} Y^{-1}_j(s)=-F_j(s) Y^{-1}_j(s).
\]
Since
\[
\f{\Id}{\Id s}\Big( Y^{-1}_1(s)Y_2(s)\Big)=Y^{-1}_1(s)(F_2(s)-F_1(s))Y_2(s)\>\>\text{ for a.e. $s\geq t_0$},
\]
one obtains the following equality (after integration and multiplication with $Y_1(t)$):
\[
Y_2(t)=Y_1(t)+Y_1(t)\int^t_{t_0} Y^{-1}_1(s)(F_2(s)-F_{1}(s))Y_2(s)\Id s. 
\]
Thus,
\begin{align}
&\left\|Y_1(t)-Y_2(t)\right\| \leq \left\|Y_1(t)\right\|\int^t_{t_0} \left\|Y^{-1}_1(s)\right\|\left\|F_2(s)-F_{1}(s)\right\|\left\|Y_2(s)\right\|\Id s.
\end{align}
The claim follows from observing that
\[
\left\|Y_j(s)\right\|\leq \mathrm{e}^{\int^t_{t_0}\left\|F_{j}(r)\right\|\Id r},\>\>\left\|Y^{-1}_j(s)\right\|\leq \mathrm{e}^{\int^t_{t_0}\left\|F_{j}(r)\right\|\Id r},
\]
which follows from the Gronwall lemma. \vspace{0.5mm}

\hfill$\blacksquare$\vspace{2mm}

Of course, similar results hold if one replaces the time interval $[t_0,\infty)$ with a finite time interval of the form $[t_0,t_1]$.\vspace{2mm}

For the second lemma, we use the notation of (\ref{for2}) and (\ref{pop1}).

\begin{Lemma}\label{opo} Let $\alpha\in\Omega^1(\IR^n, \mathscr{U}(d))$, let $V$ be a potential with
\[
0\leq V\in\mathcal{K}_{\mathrm{loc}}(\IR^n,\mathrm{Mat}(\IC^d)),
\]
and let $x,y\in\IR^n$, $t>0$, $0\leq s\leq t$. The following assertions hold: \vspace{1.2mm}

{\rm a)} One has $\mathscr{A}^{\alpha,0,*}_t= \mathscr{A}^{\alpha,0,-1}_t$ and 
\begin{align}
\left\|\mathscr{A}^{\alpha,V}_t\right\|\leq 1\>\>\text{  $\mathbb{P}^x$-a.s.} \label{blq}
\end{align}
{\rm b)} It holds that 
\[
\left\|\mathscr{A}^{\alpha,V,-1}_s\mathscr{A}^{\alpha,V}_t\right\|\leq 1\>\>\text{  $\mathbb{P}^x$-a.s.}
\]
{\rm c)} One has $\mathscr{A}^{\alpha,0,(t),*}_s= \mathscr{A}^{\alpha,0,(t),-1}_s$ and
\[
\left\|\mathscr{A}^{\alpha,V,(t)}_s\right\|\leq 1\>\>\text{ $\mathbb{P}^{x,y}$-a.s. }
\]
\end{Lemma}

{\it Proof.} Firstly, note that under these assumptions on $(\alpha,V)$, the existence of 
\[
\mathscr{A}^{\alpha,V}:[0,\infty)\times\Omega\longrightarrow\mathrm{Mat}(\IC^d)
\]
as the solution of (\ref{pop1}) with respect to $\mathbb{P}^x$, and of 
\[
\mathscr{A}^{\alpha,V,(t)}:[0,t]\times \Omega_t\longrightarrow \mathrm{Mat}(\IC^d) 
\]
as the solution of (\ref{Pop1}) with respect to $\mathbb{P}^{x,y}_t$ has been established in section \ref{zhn}. We shall prove a) and b). The proof of c) is similar to the proof of a).\vspace{1.2mm}

As we have already remarked in section \ref{zhn}, $\mathscr{A}^{\alpha,0}$ is invertible and $\mathscr{A}^{\alpha,0,-1}$ is uniquely determined by
\[
\Id \mathscr{A}^{\alpha,0,-1} =- \left( \underline{\Id} A^{\alpha,0}\right)  \mathscr{A}^{\alpha,0,-1},\>\>\mathscr{A}^{\alpha,0,-1}_0=\mathbf{1}.
\]
Noting that $A^{\alpha,0,*}=-A^{\alpha,0}$ and that $\mathscr{A}^{\alpha,0,*}$ is uniquely determined by
\[
\Id  \mathscr{A}^{\alpha,0,*} =\left( \underline{\Id}  A^{\alpha,0,*} \right)  \mathscr{A}^{\alpha,0,*},\>\>\mathscr{A}^{\alpha,0,*}_0=\mathbf{1},
\]
it follows that $\mathscr{A}^{\alpha,0}$ is unitary. \\
As in the proof of theorem \ref{aayv}, let 
\[
\tilde{\mathscr{A}}^{\alpha,V}:[0,\infty)\times\Omega\longrightarrow \mathrm{Mat}(\IC^d)
\]
be the pathwise weak solution of
\begin{align}
\f{\Id}{\Id t} \tilde{\mathscr{A}}^{\alpha,V}_t =-\tilde{\mathscr{A}}^{\alpha,V}_t  \mathscr{A}^{\alpha,0}_t V(X_t)\mathscr{A}^{\alpha,0,-1}_t,\>\>\tilde{\mathscr{A}}^{\alpha,V}_0=\mathbf{1}.\label{wqe} 
\end{align}
It follows from lemma \ref{ay11} a) that 
\[
\left\|\tilde{\mathscr{A}}^{\alpha,V}_t\right\|\leq 1\>\>\text{ $\mathbb{P}^x$-a.s.}
\]
Noting that the Stratonovic product rule implies 
\begin{align}
\mathscr{A}^{\alpha,V}_t=\tilde{\mathscr{A}}^{\alpha,V}_t\mathscr{A}^{\alpha,0}_t\>\>\text{ $\mathbb{P}^x$-a.s.,}\label{agh} 
\end{align}
inequality (\ref{blq}) follows from the fact that $\mathscr{A}^{\alpha,0}$ is unitary. \vspace{1.2mm}

b) With the notation of the proof of part a) one has 
\begin{align}
&\left\|\mathscr{A}^{\alpha,V,-1}_s\mathscr{A}^{\alpha,V}_t\right\|=\left\|\mathscr{A}^{\alpha,0}_s\tilde{\mathscr{A}}^{\alpha,V,-1}_s\tilde{\mathscr{A}}^{\alpha,V}_t\mathscr{A}^{\alpha,0,-1}_t\right\|
&\leq \left\|\tilde{\mathscr{A}}^{\alpha,V,-1}_s\tilde{\mathscr{A}}^{\alpha,V}_t\right\|.
\end{align}
Noting that for fixed $s$, the process $\tilde{\mathscr{A}}^{\alpha,V,-1}_s\tilde{\mathscr{A}}^{\alpha,V}_{\bullet}$ is the unique solution of
\begin{align}
\f{\Id}{\Id t} \left( \tilde{\mathscr{A}}^{\alpha,V,-1}_s\tilde{\mathscr{A}}^{\alpha,V}_{t}\right)  &=-\left( \tilde{\mathscr{A}}^{\alpha,V,-1}_s\tilde{\mathscr{A}}^{\alpha,V}_{t} \right)  \mathscr{A}^{\alpha,0,-1}_t V(X_t)\mathscr{A}^{\alpha,0}_t,\nn\\
\tilde{\mathscr{A}}^{\alpha,V,-1}_s\tilde{\mathscr{A}}^{\alpha,V}_{t}\mid_{t=s}&=\mathbf{1},\nn
\end{align}
the assertion follows from lemma \ref{ay11}.\vspace{0.5mm}

\hfill$\blacksquare$\vspace{2mm}




\addcontentsline{toc}{section}{Literature}

\end{document}